\documentclass[preprint,aps,12pt,preprintnumbers,eqsecnum,nofootinbib,superscriptaddress]{revtex4}
\usepackage{graphicx}
\usepackage{amssymb,amsmath}

\def\be{\begin{equation}}
\def\ee{\end{equation}}
\def\beq{\begin{eqnarray}}
\def\eeq{\end{eqnarray}}
\def\f{\frac}
\def\lf{\left}
\def\rh{\right}

\begin{document}
%
%
\title{\vspace*{0.5in} Combined Constraints on Holographic Bosonic Technicolor
\vskip 0.1in}
\author{Christopher D. Carone}\email[]{cdcaro@wm.edu}
\author{Reinard Primulando}\email[]{rprimulando@email.wm.edu}
\affiliation{Particle Theory Group, Department of Physics,
College of William and Mary, Williamsburg, VA 23187-8795}
\date{March 2010}
\begin{abstract}
We consider a model of strong electroweak symmetry breaking
in which the expectation value of an additional, possibly composite, scalar
field is responsible for the generation of fermion masses.  The dynamics of the
strongly coupled sector is defined and studied via its holographic dual, and does not
correspond to a simple, scaled-up version of QCD.  We consider the bounds from perturbative
unitarity, the $S$ parameter, and the mass of the Higgs-like scalar. We show
that the combination of these constraints leaves a relatively limited region of
parameter space viable, and suggests the qualitative features of the model that might
be probed at the LHC.
\end{abstract}
\maketitle

\section{Introduction} \label{sec:intro}

The physics of electroweak symmetry breaking (EWSB) will soon be probed directly
at the Large Hadron Collider (LHC).  One logical possibility is that the sector responsible 
for electroweak symmetry breaking will involve new, nonperturbative dynamics.  Historically, technicolor
models have represented an attempt at constructing viable theories of this
type~\cite{technicolor}.

Conventional technicolor models, however, suffer from a number of well-known problems.
As originally proposed, the technicolor sector was assumed to be a scaled-up version
of QCD, leading to estimates for the $S$ parameter that are unacceptably
large~\cite{PT}.  In addition, an extended technicolor (ETC) sector must be added to generate
the operators needed to account for Standard Model fermion masses~\cite{etc}. In many ETC models, it
is impossible to account for a heavy top quark (which requires the ETC scale to be low) and
suppress other ETC operators that contribute to flavor-changing-neutral-current (FCNC)
processes (which requires the ETC scale to be high). Viable and elegant ETC models have
been few and far between.

Developments over the last decade in the physics of higher-dimensional and conformal field
theories, however, have led to new possibilities in technicolor model
building~\cite{hong,sanz,cet1,acgr,cesh,hirayama,haba,piai,dandk,mint,kit,round,belitsky,supertc,luty,also}.
For example, the magnitude of the $S$ parameter in QCD-like technicolor theories suggests one should
not exclusively study theories that are exactly like QCD (as, for example, in Refs.~\cite{sanz}
or \cite{cet1}).  A decade or so ago, this would have been a fruitless effort. Now, the AdS/CFT
correspondence~\cite{adscft} provides a means of constructing a perturbative, five-dimensional (5D)
theory that is dual to a strongly coupled technicolor theory localized on a four-dimensional (4D)
boundary~\cite{hong,sanz,cet1,acgr,cesh,hirayama,haba,piai,dandk,mint,kit,round,belitsky}.  For some
values of the parameters that define the 5D theory, the dual theory can model a scaled-up version of
QCD.  However, for other parameter choices it does not.  In either case, observables can be computed
reliably in the 5D theory, which we can think of as defining its strongly coupled dual.  The freedom
to deviate from the QCD-like limit only presumes the validity of a gauge/gravity correspondence.  The
evidence for this is not insignificant, and includes holographic models of QCD phenomenology that
agree remarkably well with the low-energy data~\cite{adsqcd,eandw}.

The problem with fermion mass generation in the conventional ETC framework, on the
other hand, may suggest something about the form of the relevant low-energy effective
theory.  It was observed long ago that a techniquark bound state with the same quantum
numbers as a Standard Model Higgs doublet can form in ETC models in which the ETC gauge
coupling becomes strong~\cite{setc}; this bound state has Yukawa couplings to the Standard
Model fermions and may develop a vacuum expectation value, producing fermion masses.  The
low-energy effective theory, taken by itself, has no problems with FCNC effects, since
these originate from scalar-exchange diagrams that are no larger than in conventional two-Higgs
doublet models.  A significant number of phenomenological studies on such ``bosonic technicolor"
scenarios were motivated by the simplicity of this low-energy effective theory~\cite{bt1,bt2,bt3,
bt4,bt5,bt6,bt7,bt8,bt9}.  While bosonic technicolor can arise from a (fine-tuned) strongly
coupled ETC model, the low-energy effective theory is by no means linked uniquely to that
ultraviolet completion.  For example, the same effective theory can arise in a warped, 5D theory
with a Higgs field localized near the Planck brane and symmetry-breaking boundary conditions on
the bulk gauge fields~\cite{acgr}.  In this setting, the presence of a scalar in the spectrum of the
theory seems far from scandalous.  The value of working with the low-effective effective
description is that one can extract robust, low-energy predictions of the theory without
being hindered unnecessarily by one's ignorance of the physics that has decoupled in the
ultraviolet.

It is such robust predictions of the low-energy effective theory that are of interest to us
in this paper.  Using the holographic approach to define our strongly coupled sector, and
the associated freedom to deviate from the limit in which this sector is like QCD, we show
that the parameter space of the theory is nonetheless significantly constrained.  Combining the
bounds on the $S$ parameter (evaluated holographically), partial-wave unitarity
in longitudinal $W$ boson scattering (with the technirho couplings evaluated holographically)
and the bound on the mass of the light Higgs-like scalar (with a relevant chiral Lagrangian
parameter evaluated holographically), we find that there is a relatively narrow region of parameter
space in which the model is currently viable.  In deforming the theory away from its QCD-like limit,
we focus primarily on varying the ratio of the chiral symmetry breaking to the confinement scale,
as well as the amount of explicit chiral symmetry breaking originating from the current
techniquark masses.  In addition, bosonic technicolor models allow one to vary the symmetry breaking
scale associated with the strongly interacting sector, while holding the electroweak scale fixed.
Within the allowed parameter region, the ranges of observable quantities are substantially restricted,
suggesting the qualitative features of the model that may be relevant at the LHC.  The new results
presented here suggest that the simplest version of holographic bosonic technicolor may be
sufficiently constrained that upcoming collider data could soon render debates on the possible
origins of the effective theory largely irrelevant.

Our paper is organized as follows. In the next section, we review the relevant low-energy effective
theory. In Sections~III and IV, we discuss the holographic calculations of the observable quantities that we
use to constrain the parameter space of the theory.  In Section~V, we present our numerical results
and in Section~VI we summarize our conclusions.

\section{The Model} \label{sec:two}
The gauge group of the model is $G_{\rm TC} \times \textrm{SU}(3)_C \times \textrm{SU}(2)_W \times \textrm{U}(1)_Y$,
where $G_{\rm TC}$ represents the technicolor group. We will assume that $G_{TC}$ is asymptotically free and
confining. We assume two flavors of technifermions, $p$ and $m$, that transform in the $N$-dimensional representation
of $G_{\rm TC}$. In addition, these fields form a left-handed $\textrm{SU}(2)_W$ doublet and two right-handed singlets,
\be
\Upsilon_L \equiv \lf( \begin{array}{c} p \\ m \end{array} \rh)_L, \,\,\,\,\, p_R,\,\,\,\,\, m_R,
\ee
with hypercharges $Y(\Upsilon_L)=0$, $Y(p_R)=1/2$, and $Y(m_R)=-1/2$. With these assignments, the technicolor
sector is free of gauge anomalies.  With $N$ even, the SU(2) Witten anomaly is also absent.

The technicolor sector has a global $\textrm{SU}(2)_L \times \textrm{SU}(2)_R$ symmetry that is spontaneously broken
when the technifermions form a condensate
\be \label{eq:condensate}
\left< \bar pp + \bar mm \right> = \sigma_0 \,\,\,.
\ee
The electroweak gauge group of the Standard Model is a subgroup of the chiral symmetry; $\textrm{SU}(2)_W$ is isomorphic
to $\textrm{SU}(2)_L$, while $\textrm{U}(1)_Y$ is identified with the third generator of $\textrm{SU}(2)_R$. The
condensate breaks $\textrm{SU}(2)_W \times \textrm{U}(1)_Y$ to $\textrm{U}(1)_{\rm EM}$ and
generates $W$ and $Z$ masses.  However, additional physics is required to communicate this symmetry breaking
to the Standard Model fermions.  Bosonic technicolor models utilize the simplest possibility, a  scalar field $\phi$,
that transforms as an $\textrm{SU}(2)_W$ doublet with hypercharge $Y(\phi) = 1/2$. The scalar has Yukawa couplings to
both the technifermions,
\be\label{eq:techniyuk}
\mathcal L_{\phi T} =  -\bar\Upsilon_L \tilde\phi \, h_+ p_R -  \bar\Upsilon_L\phi \, h_- m_R + {\rm h.c.},
\ee
and to the ordinary fermions,
\be\label{eq:smfyuk}
\mathcal L_{\phi f} = -\bar L_L \phi h_l E_R - \bar Q_L \tilde\phi h_U U_R - \bar Q_L \phi h_D D_R + {\rm h.c.},
\ee
where $\tilde\phi = i \sigma^2 \phi^*$.  Unlike the Standard Model Higgs doublet, the $\phi$ field is assumed to
have a {\em positive} squared mass.  When the technifermions condense, Eq.~(\ref{eq:techniyuk}) produces a $\phi$
tadpole term in the scalar potential, and $\phi$ develops a vacuum expectation value, as we will see in a more
conventient parameterization below.  Standard Model fermion masses then follow from the Yukawa couplings in
Eq.~(\ref{eq:smfyuk}).

To be more explicit, we use the conventional nonlinear representation of the Goldstone bosons to construct the
electroweak chiral Lagrangian.  We define
\be\label{eq:sigdef}
\Sigma = \exp(2 i \Pi/f), \,\,\,\,\,  \Pi = \left( \begin{array}{cc} \pi^0/2 & \pi^+/\sqrt{2} \\ \pi^-/\sqrt{2} & -\pi^0/2
\end{array}\right) \, ,
\ee
where $\Pi$ represents an isotriplet of technipions, and $f$ is their decay constant. The $\Sigma$ field transforms
under $\textrm{SU}(2)_L \times \textrm{SU}(2)_R$ as
\be
\Sigma \rightarrow L\,\Sigma\, R^\dagger \,,
\ee
which dictates the form of the pion interactions.  To include the scalar doublet
consistently in the effective theory, it is convenient to use the matrix form
\be
\Phi = \lf( \begin{array}{cc} \overline{\phi^0} & \phi^+ \\ -\phi^- & \phi^0 \end{array} \rh).
\ee
The technifermion Yukawa couplings can be re-expressed as
\be
\overline{\Upsilon}_L \lf( \begin{array}{cc} \overline{\phi^0} & \phi^+ \\ -\phi^- & \phi^0 \end{array} \rh)
\lf( \begin{array}{cc} h_+ & 0 \\ 0 & h_- \end{array} \rh) \Upsilon_R \equiv \overline{\Upsilon}_L
\Phi H \Upsilon_R.
\ee
Since the underlying theory would be invariant if the combination $\Phi H$ transformed as
\be
(\Phi H) \rightarrow L \, (\Phi H)\, R^\dagger,
\ee
one may correctly include this combination in the effective theory by assuming it transforms in
this way. The lowest-order term in the electroweak chiral Lagrangian that involves $\Phi H$ is
\be \label{eq:PhiHmixing}
\mathcal L_H = c_1 4\pi f^3 \textrm{Tr}(\Phi H \Sigma^\dagger) + h.c. \; ,
\ee
where $c_1$ is an unknown, dimensionless coefficient; one expects $c_1$ to be of order one by naive dimensional
analysis~\cite{Manohar:1983md} in a QCD-like theory.  Henceforth, we assume that $h_+=h_-\equiv h$, to
simplify the parameter space of the model.

It is convenient to re-express the $\Phi$ field using a nonlinear field redefinition, similar to
Eq.~(\ref{eq:sigdef}).   Expanding about the true vacuum,
\be \label{eq:pisigma}
\Phi = \frac{\sigma+f'}{\sqrt{2}}\Sigma', \,\,\,\,\,\Sigma' = \exp(2 i \Pi'/f')\,,
\ee
where $f'$ is the vev of $\phi$ and $\Pi'$ represents its isotriplet components.    The
kinetic terms for the $\Phi$ and $\Sigma$ fields
can be written
\be \label{eq:chirallagrangian}
\mathcal L_{KE} = \frac{1}{2}\partial_\mu\sigma\partial^\mu\sigma
+\frac{f^2}{4}\textrm{Tr}(D_\mu\Sigma^\dagger D^\mu\Sigma)
+\frac{(\sigma+f')^2}{4}\textrm{Tr}(D_\mu\Sigma'^\dagger D^\mu\Sigma'),
\ee
where the covariant derivative is given by
\be
D^\mu\Sigma = \partial^\mu\Sigma-igW_a^\mu\frac{\tau^a}{2}\Sigma+ig'B^\mu\Sigma\frac{\tau^3}{2}.
\ee
In the expansion of Eq.~(\ref{eq:chirallagrangian}), there are quadratic terms that mix the gauge fields with derivatives
of a specific linear combination of the pion fields:
\be \label{eq:absorbedmixing}
\pi_a = \frac{f\,\Pi+f'\,\Pi'}{\sqrt{f^2+f'^2}}.
\ee
The mixing indicates that the components of $\pi_a$ are unphysical and can be gauged away. On the other hand, the
orthogonal linear combination,
\be \label{eq:physicalmixing}
\pi_p = \frac{-f'\,\Pi+f\,\Pi'}{\sqrt{f^2+f'^2}}\, ,
\ee
represents physical states in the low-energy theory.  The physical pion mass is determined
from Eq.~(\ref{eq:PhiHmixing}):
\be \label{eq:mpi}
m_\pi^2 = 8 \sqrt{2} \pi c_1 h \f{f}{f'} v^2  \,.
\ee
In unitary gauge, the remaining quadratic terms give the masses of $W$ and $Z$ bosons,
\be
m_W^2 = \frac{1}{4}g^2v^2,\,\,\,\,\,\,\,\,\,\,  m_Z^2=\frac{1}{4}(g^2+g'^2)v^2,
\ee
where $v$ represents the electroweak scale
\be \label{eq:vff}
v \equiv \sqrt{f^2+f'^2} = 246 \textrm{ GeV}.
\ee

In the absence of a technicolor sector,  with $f'=v$, the $\sigma$ field corresponds to the Higgs boson of the
Standard Model.  Away from this limit, the $\sigma$ field is similar to a Standard Model Higgs boson, but with
different couplings.  Expanding the third term of Eq.~(\ref{eq:chirallagrangian}), we find that
the coupling between $\sigma$ and the gauge bosons is given by
\be\label{eq:siggb}
\mathcal L_{\sigma WZ} = 2\f{f'}{v}\f{m_W^2}{v} \sigma W^{+\mu}W_\mu^- +\f{f'}{v} \f{m_Z^2}{v} \sigma Z^\mu Z_\mu \, ,
\ee
which is reduced by a factor of $f'/v$ compared to the result in the Standard Model. The couplings of the $\Phi$
field to the quarks is given by
\be \label{eq:sigmaquark}
\mathcal L_{\Phi \bar q q}= -
\bar\psi_L \Phi \lf(\begin{array}{cc} h_U&0\\ 0& V_{CKM}h_D \end{array}\rh) \psi_R + \textrm{h.c.} \ ,
\ee
where $\psi_L = (U_L, V_{CKM}D_L)$, $\psi_R = (U_R, D_R)$, $h_U = \textrm{diag}(h_u,h_c,h_t)$,
and $h_D = \textrm{diag}(h_d,h_s,h_b)$.  Using Eq.~(\ref{eq:pisigma}),  this may be written
\be
\mathcal L_{\Phi \bar q q} = - \f{\sigma + f'}{\sqrt{2}} \bar\psi_L \Sigma'
\lf(\begin{array}{cc} h_U&0\\ 0& V_{CKM} h_D \end{array}\rh) \psi_R + \textrm{h.c.}
\ee
Taking into account the leptons, the coupling of the $\sigma$ field to fermions is given by
\be\label{eq:sff}
\mathcal L_{\sigma \bar f f} = - \sum_{\textrm{fermions}} \f{v}{f'} \f{m_f}{v} \sigma \bar f f \, .
\ee
Eq.~(\ref{eq:sff}) is larger than the corresponding result in the Standard Model by a factor of $v/f'$;
this enhancement corresponds to the larger Yukawa couplings that are required when electroweak symmetry
breaking comes mostly from the strongly coupled sector.

\section{Holographic Calculations} \label{sec:three}

We model the technicolor sector using the AdS/CFT correspondence~\cite{adscft}, which allows us to numerically
evaluate the otherwise undetermined coefficients of the electroweak chiral Lagrangian, such as the parameter $c_1$ of
Eq.~(\ref{eq:PhiHmixing}).  The AdS/CFT correspondence conjectures a duality between a 5D theory
in anti-de Sitter (AdS) space and 4D conformal field theory (CFT) located on a boundary.  For theories like QCD
that are confining, the metric is a slice of AdS space:
\be
ds^2 = \f{1}{z^2}\lf(-dz^2+dx^\mu dx_\mu\rh), \ \epsilon \leq z \leq z_m.
\ee
The position in the fifth dimension $z$ corresponds to the energy scale in the 4D theory; branes at $z=\epsilon$ and
$z_m$ correspond to the ultraviolet (UV) and infrared (IR) cutoffs of the dual theory.  The AdS/CFT correspondence
dictates that operators in the boundary theory correspond to bulk fields in the 5D theory.  To be more precise, given
an operator $\mathcal O$ with the source $\phi_0(x)$, the generating functional in the 4D quantum field theory,
$W_{4D}$, is given by the classical action of the 5D theory written in terms of the boundary value of the corresponding
bulk field, $\phi(x,z)$:
\be \label{eq:correspondence}
W_{4D}[\phi_0(x)] = S_{5D}^{\rm class} |_ {\phi(x,\epsilon) = \phi_0(x)}.
\ee
In addition, the AdS/CFT correspondence identifies each global symmetry in the boundary theory with a gauge
symmetry in the 5D theory.  The 5D action that describes the technicolor sector of our model is
\be \label{eq:5Daction}
S_{5D} = \int d^5x \sqrt{g} \; \textrm{Tr} \lf\{ -\f{1}{2g_5^2}(F_R^2+F_L^2) + |DX|^2 + 3|X|^2  \rh\} \,.
\ee
The chiral symmetry of the technicolor sector corresponds to the SU(2)$_L \times $SU(2)$_R$ gauge symmetry of the
5D theory, with gauge fields $A_{L,R} = A^a_{L,R}t^a$, where the $t^a$ are generators of SU(2) with
${\rm Tr} \, t^a t^b = \delta^{ab}/2$.  The covariant derivative and field strength tensors are defined 
by $D_\mu X = \partial_\mu X - i A_{L\mu}X + iXA_{R\mu}$ and
$F_{L,R\;\mu\nu} = \partial_\mu A_{L,R\;\nu} - \partial_\nu A_{L,R\;\mu} - i \lf[A_{L,R\;\mu},A_{L,R\;\nu}\rh]$.
The scalar field $(2/z)X$ corresponds to the operator $\bar q_R q_L$, while the gauge fields $A_{L,R\;\mu}^a$
correspond to the chiral currents $\bar q_{L,R}\gamma^\mu t^a q_{L,R}$.

The equations of motion for the $X$ field may be solved, subject to the UV boundary condition
$2/\epsilon \, X(\epsilon) = m_q$:
\be\label{eq:theprof}
X(z) = \f{1}{2} (m_q z + \sigma_c z^3) \equiv \frac{1}{2} X_0(z).
\ee
The techniquark mass $m_q$ is related to the parameter $h$ by $m_q = h \, f'/\sqrt{2}$.  The coefficient
$\sigma_c$ is equal to the condensate $\sigma_0$ (defined in Eq.~(\ref{eq:condensate})) when $m_q=0$,
as can be shown by varying the action with respect to $m_q$ and then taking the chiral limit.  More generally,
$\sigma_c$ is a parameter that defines the holographic theory that we will eliminate in terms of the
technipion decay constant $f$, as we discuss below.

We work with the vector and axial vector fields $V = (A_L+A_R)/\sqrt{2}$ and $A = (A_L-A_R)/\sqrt{2}$,
respectively.  The bulk-to-boundary propagator $V(q,z)$ is defined as the solution to the transverse
equations of motion with $V_\mu(q,z)_\perp \equiv V(q,z) V_\mu(q)_\perp$ and $V(q,\epsilon)=1$, where
$\epsilon$ is the UV boundary. From Eq.~(\ref{eq:5Daction}), it follows that the bulk-to-boundary propagators satisfy
\beq
\label{eq:btbv} \partial_z\lf(\f{1}{z}\partial_z V(q,z)\rh)+\f{q^2}{z}V(q,z)&=&0 \, ,\\
\label{eq:btba} \partial_z\lf(\f{1}{z}\partial_z A(q,z)\rh)+\f{q^2}{z}A(q,z)-\f{g_5^2 X_0(z)^2}{2z^3}A(q,z)&=&0.
\eeq
In accordance with Eq.~(\ref{eq:correspondence}), the vector and axial vector two-point functions are given
holographically by~\cite{adsqcd}
\be \label{eq:pifunctions}
\Pi_V(-q^2) = \lf.\f{2}{g_5^2}\f{1}{z}\f{\partial V(q,z)}{\partial z}\rh|_{z=\epsilon}, \ \
\ \Pi_A(-q^2) = \lf.\f{2}{g_5^2}\f{1}{z}\f{\partial A(q,z)}{\partial z}\rh|_{z=\epsilon}.
\ee
where
\beq
\int d^4x \, e^{iq\cdot x} \left<J_V^{a\,\mu}(x)J_V^{b\,\nu}(0)\right> &\equiv& \delta^{ab}
\lf(\f{q^\mu q^\nu}{q^2}-g^{\mu\nu}\rh) \Pi_V(-q^2) \, ,\nonumber \\
\int d^4x \, e^{iq\cdot x} \left<J_A^{a\,\mu}(x)J_A^{b\,\nu}(0)\right> &\equiv&
\delta^{ab} \lf(\f{q^\mu q^\nu}{q^2}-g^{\mu\nu}\rh) \Pi_A(-q^2).
\eeq
Comparing $\Pi_V$ with the known perturbative result for an SU($N$) gauge theory, valid at high $q^2$,
one finds~\cite{adsqcd}
\be \label{eq:matching}
g_5^2 = \f{24\pi^2}{N} \, .
\ee
We discuss this assumption further in the section on our numerical results\footnote{The equations in
the published version of Ref.~\cite{cet1} corresponding to Eqs.~(\ref{eq:pifunctions}) and (\ref{eq:matching})
above are off by a factor of $2$.  These errors are corrected in arXiv:hep-ph/0612242 (v4).}.  With the holographic
self-energies determined, we may compute the technipion decay constant using the observation that
$\Pi_A \rightarrow -f^2$ as $q^2 \rightarrow 0$ in the chiral limit, as in Ref.~\cite{adsqcd}.  Since
we treat $f$ as an input parameter, this computation may be inverted to solve for the parameter
$\sigma_c$ defined in Eq.~(\ref{eq:theprof}).  The holographic model of the technicolor sector is
then determined by three free parameters: $h$, $f$, and $z_m$.

The IR cutoff $z_m$, however, may be eliminated in terms of a single physical observable, the technirho
mass.  The technirho corresponds to the lowest normalizable mode of the 5D vector field.  The technirho
wave function $\psi_\rho(z)$ satisfies the same equation of motion as the the bulk-to-boundary propagator
in Eq.~(\ref{eq:btbv}), but with different boundary conditions: $\psi_\rho(\epsilon)=0$ and
$\partial_z\psi_\rho(z_m)=0$.  These boundary conditions are satisfied when $q^2=m_\rho^2$ (or
the squared mass of any higher vector mode). The vector equation of motion may be solved analytically,
and one finds that $z_m$ is determined by
\begin{equation}
J_0(m_\rho \, z_m) =0  \,\,\,.
\end{equation}
Hence, for fixed values of $h$, the $f$-$m_\rho$ plane provides a convenient visual representation
of the parameter space of the model.

For our subsequent analysis, we will need the coupling of the technirho to the physical pion states.
Couplings between modes may be obtained by substituting properly normalized wave functions into the
appropriate interaction terms of the 5D theory, and then integrating over the extra dimension.
Requiring that the 4D kinetic terms of the technirho are canonical gives us the normalization
condition
\begin{equation}
\int(dz/z)\psi_\rho(z)^2=1 \, .
\end{equation}
For the pions, the situation is slightly more complicated. There is an isotriplet component $\pi$
of the $X$ field, $X=X_0\exp(2 \, i\,  \pi^a \, t^a)$; the longitudinal component of the axial vector
field, $\varphi$ is also an isotriplet, $A_M = A_{M\perp} + \partial_M\varphi$.  These fields satisfy
the coupled equations of motion
\beq \label{eq:pioneom}
\partial_z \lf(\f{1}{z}\partial_z\varphi^a\rh)+\f{g_5^2X_0^2}{\sqrt{2}z^3}
\lf(\pi^a-\f{\varphi^a}{\sqrt{2}}\rh)&=&0 \, ,\nonumber \\
-\sqrt{2}q^2\partial_z\phi^a+\f{g_5^2X_0^2}{z^2}\partial_z\pi^a&=&0 \, .
\eeq
The technipion state $\Pi$ corresponds to an eigensolution of the form $\pi(q,z) = \pi(z)\Pi(q)$ and
$\varphi(q,z)=\varphi(z)\Pi(q)$, subject to the boundary conditions
$\varphi'(z_m)=\varphi(\epsilon)=\pi(\epsilon)=0$~\cite{adsqcd}.  Again, requiring a canonical
4D kinetic term for the $\Pi$ field gives the desired normalization condition
\be
\int dz \; \lf(\f{\varphi'(z)^2}{g_5^2\;z}+\f{X_0(z)^2\lf(\pi(z)-\varphi(z)/\sqrt{2}\rh)^2}{z^3}\rh) = 1.
\ee
The $\rho \,\Pi \,\Pi$ coupling originates from  $V\varphi\varphi$, $V\varphi\pi$ and
$V\pi\pi$ interactions in the 5D theory, evaluated on the lowest modes:
\be \label{eq:grhoPiPi}
g_{\rho \,\Pi\,\Pi} = \f{g_5}{\sqrt{2}}\int dz \; \psi_\rho(z) \lf(\f{\varphi'(z)^2}{g_5^2\;z}
+\f{X_0(z)^2\lf(\pi(z)-\varphi(z)/\sqrt{2}\rh)^2}{z^3}\rh).
\ee
This result is not quite what we need since we have not taken into account that physical pion states
in the bosonic version of the theory involve mixing between the $\Pi$ and $\Pi'$ fields.   The mass of
the $\Pi$ field that follows from Eq.~(\ref{eq:pioneom}) corresponds to the $\Pi^2$ part of the chiral
Lagrangian term in Eq.~(\ref{eq:PhiHmixing}), allowing us to fix the coefficient $c_1$.  It follows that the
physical pion mass and the $\Pi$ mass are related by
\be \label{eq:mPi}
m_\Pi^2 = m_\pi^2 \frac{f'^2}{v^2}
\ee
where $m_\Pi^2$ is the $q^2$ eigenvalue of Eq.~(\ref{eq:pioneom}).

Following from the 5D Lagrangian, the generic interaction between the technirho and the $\pi_a$
and/or $\pi_p$ fields is of the form
\be \label{eq:rhopion}
\mathcal L_{\rho XY} = ig_{\rho XY} \rho_0^\mu\lf[(\partial_\mu X^+)Y^- - Y^+(\partial_\mu X^-)\rh],
\ee
where $X$ and $Y$ are either a physical or absorbed pion. Taking into account the mixing
in Eqs.~(\ref{eq:absorbedmixing}-\ref{eq:physicalmixing}), it follows from Eq.~(\ref{eq:grhoPiPi}) that
\beq \label{eq:gcoupling}
g_{\rho\pi_a\pi_a} &=& \f{f^2}{v^2} g_{\rho\Pi\Pi} \, , \nonumber \\ g_{\rho\pi_a\pi_p}
=g_{\rho\pi_p\pi_a} &=& \f{ff'}{v^2} g_{\rho\Pi\Pi} \, , \nonumber \\ g_{\rho\pi_p\pi_p}
&=& \f{f'^2}{v^2} g_{\rho\Pi\Pi} \, .
\eeq
For the masses of the technirho of interest to us later, the $\pi_a$ couplings will accurately
describe the coupling of the technirho to longitudinal $W$ bosons via the Goldstone boson equivalence
theorem.

\section{Constraints on the model}\label{sec:four}
\subsection{$S$ Parameter}

The size of the $S$ parameter represents a significant challenge for most technicolor theories~\cite{PT}.
Electroweak precision tests favor a value smaller than $0.09$~\cite{Amsler:2008zzb}; we use this
fact to exclude regions of the $f$-$m_\rho$ plane.  The $S$ parameter may be defined in terms of
the self-energies $\Pi_V$ and $\Pi_A$, which are computed holographically via
Eq.~(\ref{eq:pifunctions}):
\be\label{eq:sdefine}
S = 4\pi \f{d}{dq^2} \lf.\lf(\Pi_V(-q^2)-\Pi_A(-q^2)\rh)\rh|_{q^2\rightarrow 0} \, .
\ee
Note that the dependence of the self energies on the ultraviolet cut off, $1/\epsilon$, cancels
between the two terms in Eq.~(\ref{eq:sdefine})\footnote{Given that we extract $f$ from $\Pi_A$
alone, is it worth noting that the $\ln\epsilon$ dependence in this self-energy vanishes for $q^2=0$
in the chiral limit.  However, for $m_q \neq 0$, there is a divergence proportional to
$m_q^2 \ln \epsilon$ in $\Pi_A(0)$ that we subtract.  In the language of chiral perturbation theory, this
is equivalent to adding a counterterm whose unknown finite part is of order $m_q^2$.  For the techniquark
masses we consider, this is a negligible correction.}. It was shown in Ref.~\cite{cet1}, that the
value of the $S$-parameter may be reduced by decreasing $f$ or increasing $m_\rho$. (The same effect has
been described in a different context in Ref.~\cite{acp}.) In the first case, one approaches the limit where
electroweak symmetry breaking is accomplished almost entirely by the $\phi$ field, so the
presence of technihadronic resonances is irrelevant; in the second case, the technihadronic
resonances are decoupled, which also reduces the result for fixed $f$.

\subsection{Unitarity}
\begin{figure}[t]
    \centering
        \includegraphics[width=75mm,angle=0]{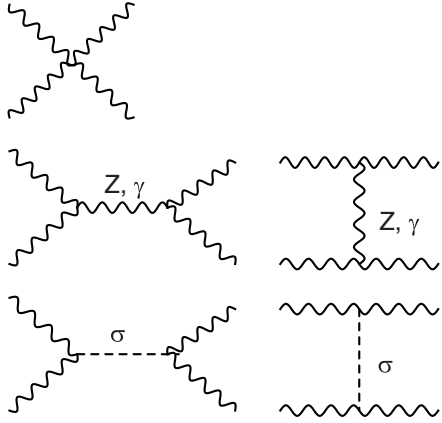}
    \caption{Gauge and $\sigma$-boson contributions to $WW\rightarrow WW$ scattering.}
    \label{fig:Unitary Higgs}
\end{figure}
\begin{figure}[t]
    \centering
        \includegraphics[width=75mm,angle=0]{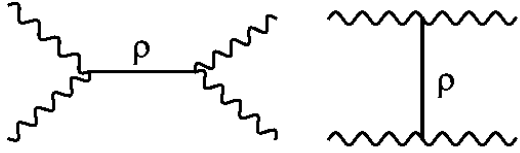}
    \caption{Technirho contributions to $WW\rightarrow WW$ scattering.}
    \label{fig:Unitary Rho}
\end{figure}

In the Standard Model, unitarity of the $W^+W^- \rightarrow W^+W^-$ scattering amplitude can
be used to obtain a constraint on the Higgs boson mass~\cite{Lee:1977eg}. In the present model,
the $\sigma$ field is analogous to the Standard Model Higgs boson, but its coupling to $W^+W^-$
is reduced by a factor of $f'/v$, as we saw in Eq.~(\ref{eq:siggb}).  The Feynman diagrams
involving gauge fields and the $\sigma$ boson are shown in Fig.~\ref{fig:Unitary Higgs}.  The
corresponding amplitude is given at leading order by
\be
\mathcal M_{\rm gauge} + \mathcal M_{\sigma} = \f{1}{v^2}\lf(s+t\rh)
-\f{1}{v^2}\lf(\f{f'}{v}\rh)^2\lf(\f{s^2}{s-m_\sigma^2}+\f{t^2}{t-m_\sigma^2}\rh).
\ee
for momenta large compared to $m_W$.  In addition, the scattering amplitude also receives important
contributions from diagrams involving technirho exchanges, shown in Fig.~\ref{fig:Unitary Rho}.  To evaluate
these, we use the Goldstone boson equivalence theorem and compute the technirho-exchange
contributions to $\pi_a^+ \pi_a^- \rightarrow \pi_a^+ \pi_a^-$, where $\pi_a$ is the linear
combination of isotriplet fields that would be absent in unitary gauge.  This will give the
longitudinal $W$ boson scattering amplitude accurately for external momenta large compared
to the $W$ boson mass, a criterion that will always be satisfied in the regions of parameter
space that are of interest to us.  The $\rho\pi_a\pi_a$ coupling is given by Eqs.~(\ref{eq:rhopion})
and (\ref{eq:gcoupling}), with $g_{\rho\Pi\Pi}$ computed holographically using Eq.~(\ref{eq:grhoPiPi}).
Thus, we find the technirho contribution to the scattering amplitude
\be
\mathcal M_\rho = g_{\rho \pi_a\pi_a}^2 \lf(\f{s+2t}{s-m_\rho^2}+\f{2s+t}{t-m_\rho^2}\rh),
\ee
and the total amplitude
\be
\mathcal M = \mathcal M_{\rm gauge} + \mathcal M_{\sigma} + \mathcal M_\rho.
\ee
Note that the total amplitude is gauge invariant, as is $\mathcal M_\rho$ separately.

The most significant constraint from unitarity can be obtained by considering the $J=0$
partial wave
\be
a_0(s) = \f{1}{16\pi s} \int_{-s}^0 \mathcal M \, dt.
\ee
Following Ref.~\cite{hhg} we require $|\textrm{Re } a_0(s)| \leq 1/2$, over the range
of energies in which our holographic calculation is valid.  Based on what is known from
holographic models of QCD, the holographic construction is trustworthy up to the mass of
the lowest vector and axial-vector resonances, but becomes increasingly less accurate
when properties of heavier hadronic resonances are considered.  Thus, we take the mass
scale of the second vector resonance ({\em i.e.}, the first excited state of the technirho)
as a cutoff for our effective theory.  If unitarity is violated above this scale, no conclusion
can be drawn because the calculational framework is suspect. If unitarity is violated below this
scale, the effective theory is excluded in its minimal form.  For the $\sigma$ boson taken as
light as possible, the technirho cannot be made arbitrarily heavy without violating this
constraint.  However, the lower bound on the technirho mass is relaxed when $f$ is made
small since the model mimics the Standard Model in this limit.

\subsection{Higgs Mass}

As mentioned in the previous section, the $\sigma$ field is similar to the Standard Model
Higgs boson, but with modified couplings.  If light enough, the $\sigma$ boson would have
been produced at LEP via the Higgstrahlung process $e^+e^- \rightarrow Z^* \rightarrow \sigma Z$.
In the region of parameter space left viable after the consideration of the unitarity and
$S$ parameter bounds (discussed in the next section), the $\sigma-Z$ coupling is not less
than $\sim 90$\% of its Standard Model value.  In this case, the LEP bound is modified in
accordance with Fig.~10 of Ref.~\cite{Barate:2003sz}, and differs negligibly from the Standard
Model result.  Note that the partial decay widths to two fermions are slightly enhanced
while the partial decay widths to two fermions and one gauge boson (via an off-shell gauge boson)
are slightly suppressed.  Hence, the branching fraction to the primary decay channels at LEP,
namely $\bar b b$ and $\bar \tau \tau$, will remain practically unaffected.  Thus, we will
apply the bound $m_\sigma \geq 114.4 \textrm{ GeV}$ to constrain the parameter space of the
model.

The possible perturbative interactions of the $\phi$ field allow us to construct a potential
for $\sigma$.  Following the conventions of Ref.~\cite{bt3},
\be \label{eq:V}
V(\sigma_s) = \f{M^2}{2}\sigma_s^2 + \f{\lambda}{8}\sigma_s^4 -
\f{1}{64\pi^2}\lf[3h_t^4 + 2Nh^4\rh]\sigma_s^4\ln\lf(\f{\sigma_s^2}{\mu^2}\rh)
-8\sqrt{2}\pi c_1 f^3 h\sigma_s \, ,
\ee
where $M^2 \geq 0$ and $\sigma_s = \sigma + f'$.  The third term represents the one-loop radiative corrections
from the top quark and the techniquarks, though only the former is substantial. All other radiative corrections
can be neglected for the values of the couplings that are relevant in the next section.  In order to remove
the dependence on the renormalization scale $\mu$, we define a renormalized coupling
$\lambda_r \equiv 1/3 \, V''''(f')$, where primes refer to derivatives with respect to $\sigma_s$.  It 
is convenient for us to work with a redefined coupling $\tilde{\lambda}$, where
\be \label{eq:Mlambda}
\tilde \lambda \equiv \lambda_r + \f{11}{24\pi^2}\lf[ 3h_t^4+2Nh^4 \rh] \, .
\ee
Since the $\sigma$ field has no vacuum expectation value, $V'(f')=0$, from which it follows that
\be \label{eq:novevrelation}
M^2f' + \f{1}{2}\tilde \lambda f'^3 = 8\sqrt{2}c_1\pi f^3 h \, .
\ee
The mass of $\sigma$ is given by $V''(f')$:
\be
m_\sigma^2 = M^2 + \lf( \f{3}{2}\tilde \lambda - \f{1}{8\pi^2}\lf[3h_t^4+2Nh^4\rh] \rh)f'^2 \, .
\ee
We can eliminate $\tilde \lambda$ using Eq.~(\ref{eq:novevrelation}), as well as the chiral Lagrangian
parameter $c_1$ using Eqs.~(\ref{eq:mpi}) and (\ref{eq:Mlambda}):
\be
m_\sigma^2 = 3 m_\pi^2 \f{f^2}{v^2} - \f{1}{8\pi^2}\lf[3h_t^4+2Nh^4\rh]f'^2 - 2 M^2 \, .
\ee
Since the last term is no smaller than zero in the models of interest\footnote{If $M^2<0$, EWSB 
occurs whether or not there is a technicolor condensate, and the model is different in spirit (and 
arguably less interesting) then the model we consider here.  We do not discuss this possibility
further.}, we conclude that
\be\label{eq:msigbnd}
m_\sigma^2 \leq 3 m_\pi^2 \f{f^2}{v^2} - \f{1}{8\pi^2}\lf[3h_t^4+2Nh^4\rh]f'^2 \, .
\ee
The physical pion mass $m_\pi$ is computed holographically following the discussion
of Sec.~\ref{sec:three}.  For any region of the $f$-$m_\rho$ plane where the right-hand
side of Eq.~(\ref{eq:msigbnd}) is less than the LEP bound, the $\sigma$ mass can never
be any larger, for any positive $M^2$.

\section{Numerical Results}\label{sec:five}
\subsection{Allowed Regions}

In this subsection, we present our results for the allowed region of the model's parameter space.
We first assume $h=0.01$ and that the value of $g_5$ is the same as in an SU($4$) technicolor sector.
For the unitarity calculation, we fix the $\sigma$ mass at the LEP bound, $114.4$~GeV; taking the
$\sigma$ mass higher only makes the unitarity bound on the $\rho$ mass stronger.  The excluded regions
are plotted on the $f/v$ versus $m_\rho$ plane.  We later consider how the excluded regions
change as $h$ and $g_5$ are varied.

\begin{figure}
    \centering
        \includegraphics[width=15cm,angle=0]{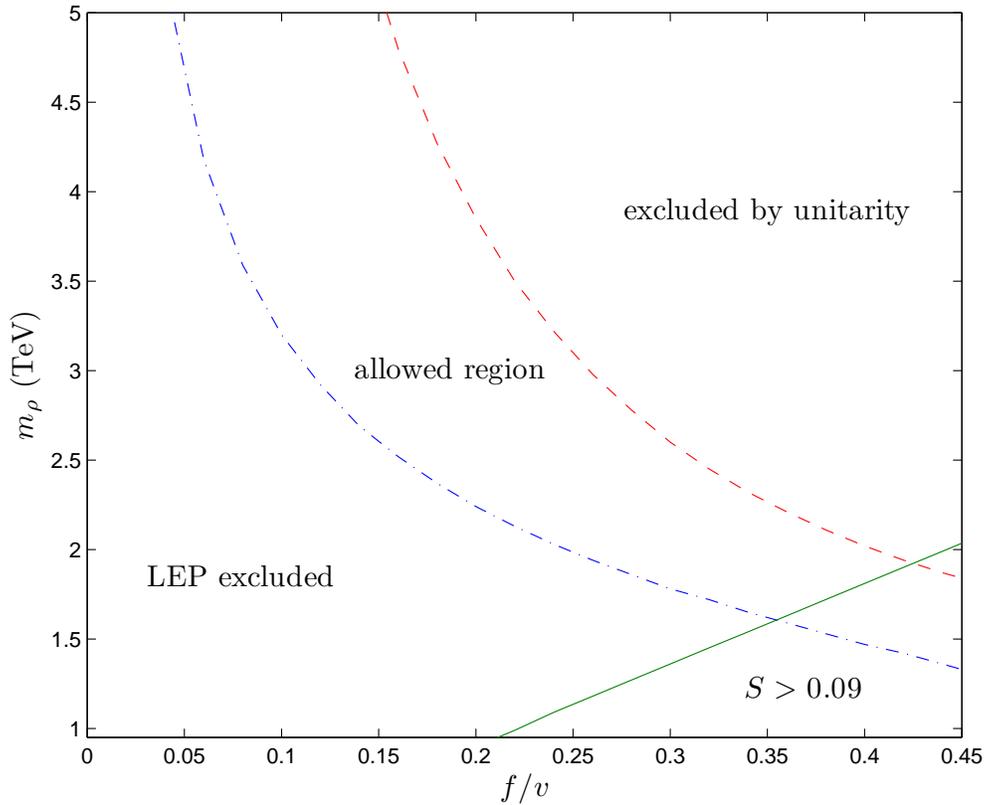}
    \caption{The allowed region for $h = 0.01$.}
    \label{fig:Exclusionh001}
\end{figure}

Our results are presented in Fig.~\ref{fig:Exclusionh001}.  The bound from the $S$ parameter
eliminates the portion of the plot with large $f$ and small technirho masses.   In this region,
the mass scale of the technihadrons is low and electroweak symmetry breaking is primarily a
consequence of technicolor dynamics; one would expect this to correspond to a problematic value
of the $S$ parameter.  On the other hand, the unitarity constraint excludes the region with
large $f$ and large technirho masses.  Here the theory is more technicolor-like, and the technirho
has a greater impact than the $\sigma$ boson in unitarizing the theory.  Finally, for small
values of $f$ and small technirho masses, there is no value of $M^2 \geq 0$ for which the $\sigma$ boson
mass is as large as the LEP bound.  The allowed region is represented by the narrow band in
the central region of Fig.~\ref{fig:Exclusionh001}.  The intersection of the boundaries from the
$S$ parameter  and $\sigma$ mass bounds gives us a lower bound on the technirho mass:
\be
m_\rho \geq 1.6 \textrm{ TeV}\, , \,\,\,\,\, (h=0.01) \,.
\ee

\begin{figure}
    \centering
        \includegraphics[width=15cm,angle=0]{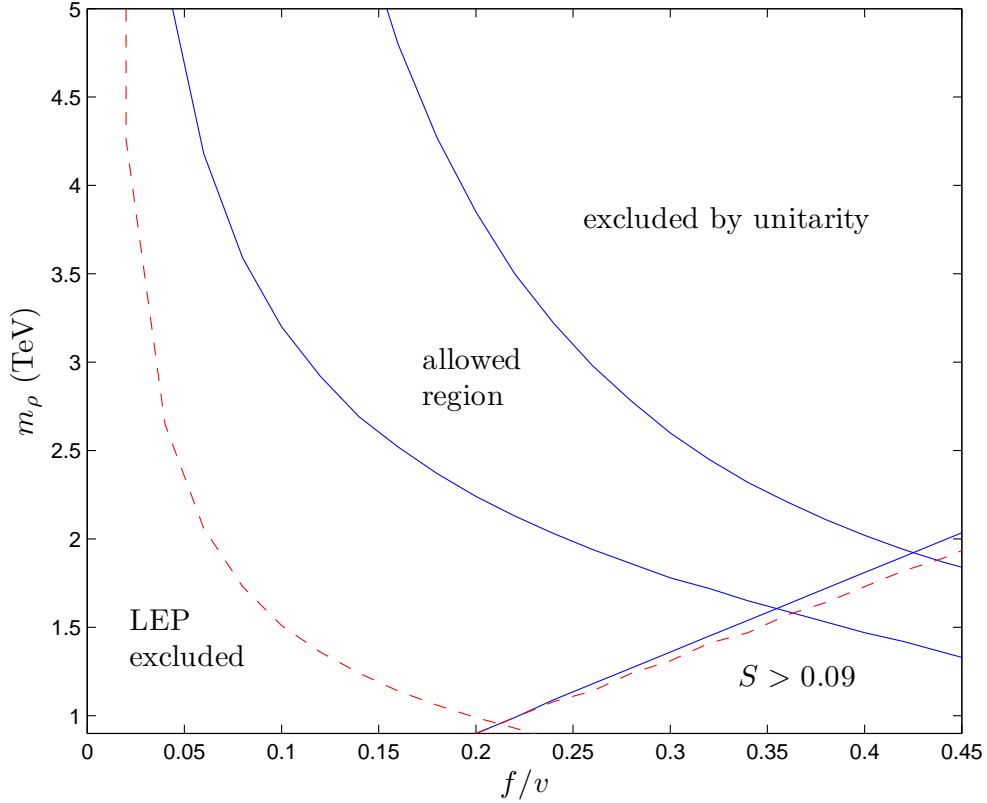}
    \caption{The allowed region as the value of $h$ is varied.  The solid (dashed) lines
correspond to $h=0.01$ $(0.05)$.  The unitarity exclusion lines for $h=0.01$ and $h=0.05$ coincide
and are represented by a single solid line.}
    \label{fig:varh}
\end{figure}

The allowed region and the lower bound on technirho mass can change if we vary
the assumed values of $h$ and $g_5$.  In Fig.~\ref{fig:varh}, we show the effect
of increasing the techniquark Yukawa coupling $h$ from $0.01$ to $0.05$.  While
the unitarity and $S$ parameter exclusion regions are only slightly affected,
the boundary of the LEP-excluded region is shifted noticeably.  For a fixed
point in the $f/v$-$m_\rho$ plane, taking $h$ larger ({\em i.e.}, making the
techiquarks heavier) increases the minimium possible mass of the $\sigma$ boson,
by increasing the technipion mass in the first term of Eq.~(\ref{eq:msigbnd}).
Hence, the exclusion line shifts to lower values of the technirho mass, where
$m_\pi$ is again reduced. A consequence of the enlarged allowed region is that
the absolute lower bound on the technirho mass is relaxed:
\be
m_\rho \geq 960 \textrm{ GeV}\, , \,\,\,\,\, (h=0.05)\,.
\ee
One may reasonably ask what happens to the allowed parameter space as one
varies $h$ further.  For larger values of $h$, $m_q/\sigma_c^{1/3}$ quickly becomes
of order one: the assumption of approximate chiral symmetry is lost and
the predictions of the holographic theory are no longer trustworthy.  If one requires
$m_q/\sigma_c^{1/3} <1/3$ everywhere in the allowed region, then the largest possible
techniquark Yukawa coupling is $h=0.18$, and one finds $m_\rho \geq 630$~GeV.
On the other hand, if $h$ is taken smaller that $0.01$, the exclusion line from
the LEP bound moves toward larger $m_\rho$, while the others do not change appreciably.
For $h<1.0\times 10^{-3}$, no allowed region remains.

\begin{figure}
    \centering
        \includegraphics[width=15cm,angle=0]{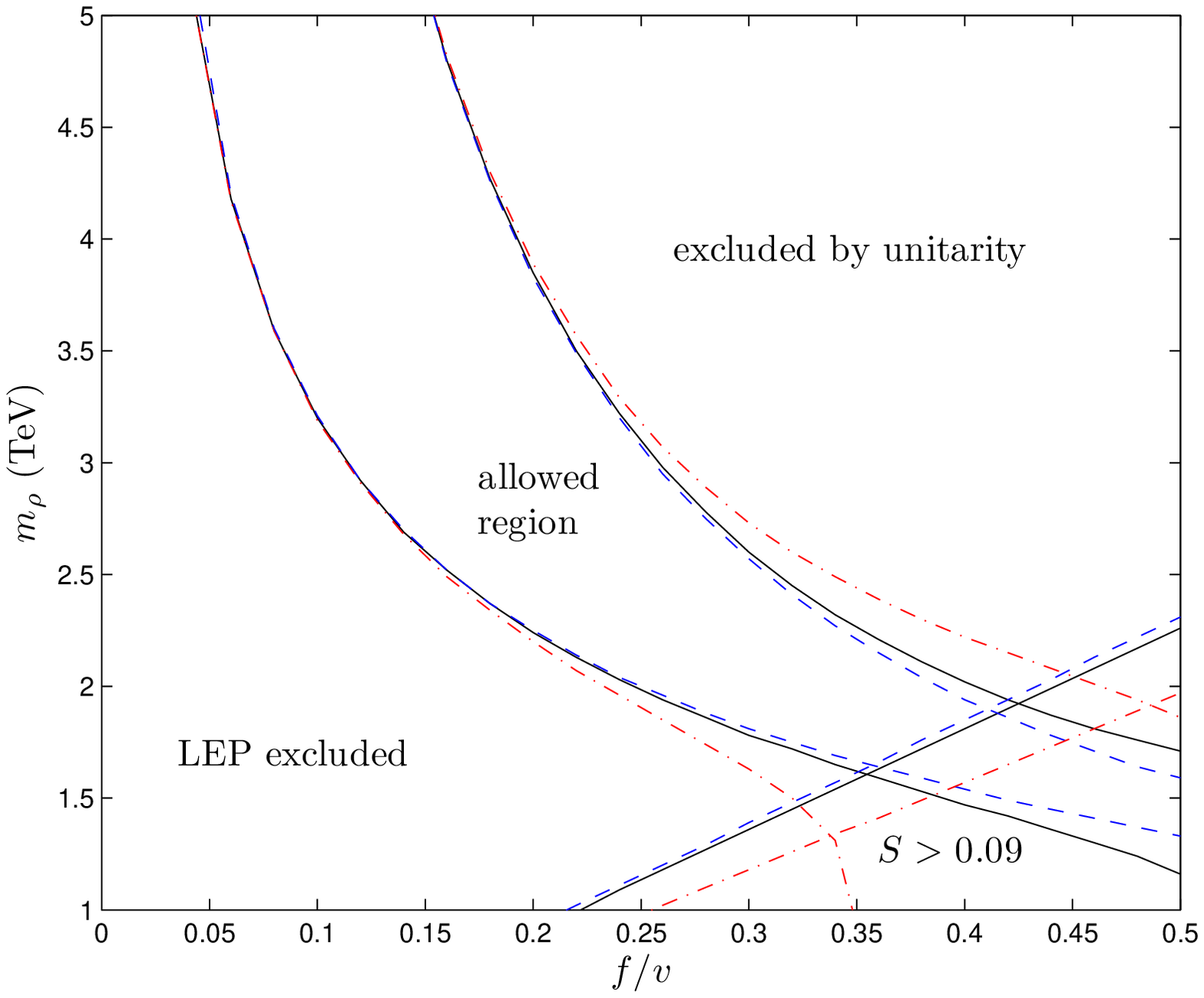}
    \caption{The allowed region as the value of $g_5$ is varied. The solid, dashed and dot-dashed lines
correspond to $r=1$, $0.5$ and $2.0$ respectively, where $g_5 =  r \sqrt{24\pi^2/N}$ with $N=4$.}
\label{fig:varg5}
\end{figure}

In constructing the holographic model of the technicolor sector, the 5D gauge coupling was chosen so
that current-current correlators would have the same high-$q^2$ behavior as in an SU($N$) gauge theory.
The same approach is used in successful holographic models of QCD~\cite{adsqcd}, where one
knows with certainty that the gauge theory of interest is SU(3).  In our case, this choice simply defines
the class of models that we choose to study, and allows us to make definite phenomenological
predictions.  While predictivity requires us to make some well-motivated choice for $g_5$, it
is still useful study whether our predictions are sensitive to the precise
value chosen.  To do so, we allow $g_5$ to vary by half and twice the value given in
Eq.~(\ref{eq:matching}).  The results are shown in Fig.~\ref{fig:varg5}.  One can see that
the qualitative changes in the shape of the allowed region are not particularly dramatic.

\subsection{Technirho Decays}

Since the allowed parameter space of Fig.~\ref{fig:Exclusionh001} that is within the reach of the LHC
is limited, it is interesting to see whether observable quantities vary appreciably within this region.  We
focus on the $\rho$ and technipion masses, as well as the dominant $\rho$ branching fractions.  The technirho
couples most strongly to the technipion field $\Pi$, which is partly $\pi_p$ and $\pi_a$.
Using the fact that the technipion mass is considerably smaller than technirho mass, the
decay to absorbed technipions is equal to the decay to longitudinal $W$ bosons by
the Goldstone boson equivalence theorem. The interaction Lagrangian is defined
in Eqs.~(\ref{eq:rhopion}) and (\ref{eq:gcoupling}), with the coupling $g_{\rho\,\Pi\,\Pi}$
calculated holographically using Eq.~(\ref{eq:grhoPiPi}).  The associated decay widths are
given by~\cite{cet1}
\beq
\Gamma_{\pi_p\pi_p} &=& \f{1}{48\pi}m_\rho g^2_{\rho\pi_p\pi_p}\lf(1-4\f{m_\pi^2}{m_\rho^2}\rh)^{3/2},
\nonumber \\
\Gamma_{W_L W_L} &=& \f{1}{48\pi}m_\rho g^2_{\rho\pi_a\pi_a}\lf(1-4\f{m_W^2}{m_\rho^2}\rh)^{3/2},
\nonumber \\
\Gamma_{W_L^{\pm}\pi_p^{\mp}}  &
=& \f{1}{48\pi}m_\rho g^2_{\rho\pi_p\pi_a}\lf(1+\f{m_\pi^4}{m_\rho^4}
+\f{m_W^4}{m_\rho^4}-2\f{m_W^2}{m_\rho^2}-2\f{m_\pi^2}{m_\rho^2}-2\f{m_\pi^2m_W^2}{m_\rho^4}\rh)^{3/2}.
\eeq
There are many subleading decay modes that one could also consider.  Each could be evaluated by
a holographic calculation, in some cases requiring the modification of the 5D theory to include
additional fields.  A complete analysis goes beyond the scope of the present work.  However, we
will consider the decay to dileptons here, since this represents a particularly clean
channel for searches at the LHC.  This decay proceeds via the vector-meson dominance couplings of the
technirho to the photon and the $Z$.  In the 5D theory, the gauge fields of a weakly gauged subgroup
of the global chiral symmetry of the boundary theory appear as coefficients of the non-normalizable
modes of the bulk gauge fields~\cite{ss}.  Substituting these into the 5D Lagrangian and integrating over the
extra dimension yields the desired couplings:
\be
\mathcal L = -\frac{m_\rho^2}{f_\rho} \left[ e A_\mu + \frac{e}{2 s_\theta c_\theta} (c_\theta^2-s_\theta^2)
Z_\mu\right]
\rho^\mu  \, .
\ee
Here $s_\theta$ ($c_\theta$) represents the sine (cosine) of the weak mixing angle.  The technirho decay
constant is given by
\be
f_\rho = \frac{1}{2} \, g_5 \, (m_\rho z_m)\, J_1(m_\rho\, z_m) \,\,.
\ee
Since the product $m_\rho \,z_m$ is fixed for the lowest vector resonance, one finds
$f_\rho \approx 4.8$.  The decay width to a single flavor of lepton is then straightforward
to compute:
\beq
\Gamma_{e^+ e^-} &=& \frac{4\pi\alpha_{EM}^2}{3f_\rho^2}
m_\rho \nonumber \\ && \; \times\left[\left( Q_e +c_{V,e}\frac{c_\theta^2-s_\theta^2}
{4s_\theta^2 c_\theta^2}\frac{m_\rho^2}{m_\rho^2-m_Z^2}\right)^2+\left( c_{A,e} \frac{c_\theta^2-s_\theta^2}
{4s_\theta^2 c_\theta^2}\frac{m_\rho^2}{m_\rho^2-m_Z^2}\right)^2\right].
\eeq
Here $Q_e = -1$, $c_{V,e} = -1/2 + 2s_\theta^2$, and $c_{A,e} = -1/2$.  The total decay width is obtained
by summing the partial widths for all decay channels.  Ignoring some of the possible subleading modes
only provides small corrections to the branching fractions that we consider here.

\begin{table}
    \centering
        \begin{tabular}{|c|c|c|c|c|c|c|c|c|}
        \hline
        No. & $m_\rho$ (TeV) & $f/v$ & $m_\pi/m_\rho$ & $\Gamma_\rho/m_\rho$ & $\textrm{BR}_{WW}$ (\%)
& $\textrm{BR}_{W\pi}$ (\%)& $\textrm{BR}_{\pi\pi}$ (\%)& $\textrm{BR}_{e^+e^-}$ (\%)\\
        \hline
        1& 1.59& 0.36& 0.13& 0.23& 1.8& 23.3& 74.9& $5\times 10^{-3}$\\
        2& 1.90& 0.36& 0.12& 0.24& 1.8& 23.3& 74.9& $5\times 10^{-3}$\\
        3& 2.21& 0.36& 0.12& 0.24& 1.8& 23.3& 74.8& $5\times 10^{-3}$\\
        4& 2.21& 0.29& 0.13& 0.25& 0.78& 16.1& 83.1& $5\times 10^{-3}$\\
        5& 2.21& 0.22& 0.15& 0.24& 0.27& 9.8& 89.9& $5\times 10^{-3}$\\
        6& 2.90& 0.22& 0.15& 0.24& 0.27& 9.8& 89.8& $5\times 10^{-3}$\\
        7& 3.50& 0.22& 0.15& 0.25& 0.27& 9.8& 89.8& $5\times 10^{-3}$\\
        8& 3.50& 0.16& 0.18& 0.23& 0.08& 5.5& 94.4& $5\times 10^{-3}$\\
        9& 3.50& 0.11& 0.22& 0.20& 0.01& 2.4& 97.6& $6\times 10^{-3}$\\
        10& 5.00& 0.15& 0.18& 0.23& 0.06& 4.9& 95.0& $5\times 10^{-3}$\\
        11& 5.00& 0.10& 0.22& 0.21& 0.01& 2.4& 97.6& $6\times 10^{-3}$\\
        12& 5.00& 0.05& 0.31& 0.14& $0.001$& 0.76& 99.1& $8\times 10^{-3}$\\
        \hline
        \end{tabular}
    \caption{Technirho decay table for $h=0.01$}
    \label{tab:TechnirhoDecay}
\end{table}

\begin{figure}
    \centering
        \includegraphics[width=15cm,angle=0]{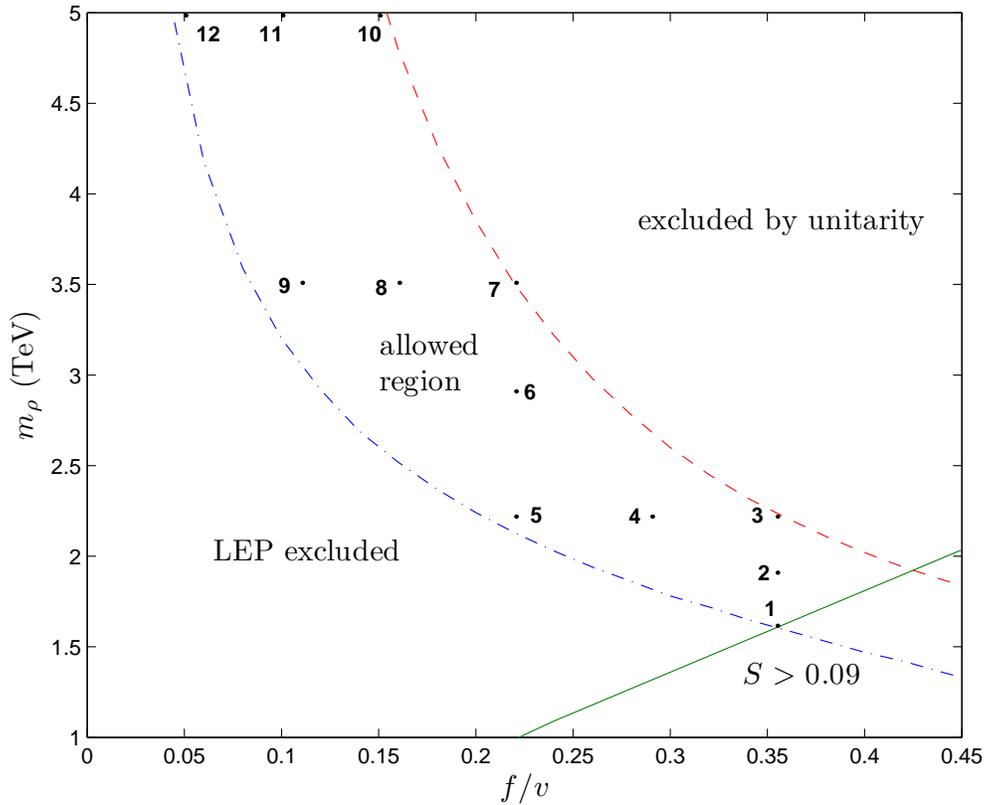}
    \caption{Twelve sample points considered in Table \ref{tab:TechnirhoDecay}.  The allowed
region assumes $h = 0.01$.}
    \label{fig:h001points}
\end{figure}

Table~\ref{tab:TechnirhoDecay} presents our results for the case $h=0.01$, over a set of
sample points within the allowed region of the model's parameter space.  The location of the
sample points is shown in Fig.~\ref{fig:h001points}.  From the table, we see that the total
decay width depends almost solely on $m_\rho$ for $m_\rho<5$~TeV (we don't consider
larger masses, which are not likely to be within the reach of the LHC).  The branching fractions,
on the other hand, depend mostly on $f/v$. As $f/v$ becomes smaller, the branching fraction
of the technirho to two physical pions increases, since the other dominant decay channels,
$W_L\pi$ and $W_L W_L$, are suppressed by factors of $f/v$ and $f^2/v^2$, respectively.  Everywhere
in the allowed parameter space the decay mode to $\pi_p \pi_p$ is dominant.  The branching fraction
to dileptons varies only between $5\times 10^{-5}$ and $8 \times 10^{-5}$, always significantly suppressed
compared to the leading modes.  We have estimated that detection of the technirho via its decays to
dileptons at the LHC would be feasible only if the dominant modes to technipions were kinematically
forbidden.  However, this favorable situation only occurs in regions of parameter space that are
excluded by the bounds we have considered.

\section{Conclusions}\label{sec:concl}

We have shown how the combined constraints from the $S$ parameter, partial-wave unitarity
and searches for a light Higgs-like scalar, meaningfully limit the viable parameter space
of a simple holographic bosonic technicolor model.  The parameter space of the model itself indicates
an important difference between this model and conventional, QCD-like technicolor theories:
different points in the $f$-$m_\rho$ plane have different ratios of the chiral symmetry breaking
scale to the confinement scale.  In QCD-like technicolor, this ratio is fixed.  The $S$ parameter
eliminates the region where $f$ is large and $m_\rho$ is small.  Here, technicolor dynamics
is the primary agent responsible for electroweak symmetry breaking.  Perturbative unitarity
eliminates the region where $f$ is large and $m_\rho$ is large.  Here, the technirho is
more important than the Higgs-like scalar in unitarizing the theory.  Finally, the LEP bound
on the mass of the Higgs-like scale eliminates the region where both $m_\rho$ and $f$ are small.
Below $m_\rho < 5$~TeV, a limited region of allowed parameter space remains.  We pointed
out a number of physical quantities (for example, the ratio of the physical pion to technirho mass and
the technirho branching fraction to two pions) that do not vary strongly within this region.  We
also studied how the allowed region changes as the techniquark mass and the 5D gauge coupling are
varied.

In the near future, data from the LHC may make it possible to rule out this type of model, without
recourse to philosophical or aesthetic arguments.  For example, something as simple as a tighter lower
bound on the neutral scalar mass could substantially squeeze or eliminate the allowed band in
Fig.~\ref{fig:Exclusionh001}.  As another example, we have found that within the allowed region, the
branching fraction of the technirho to two physical pions varies between 75-100\%, suggesting a channel
for future collider studies.  Other decay modes that we have not considered may be of value in excluding
additional parameter space, but these require additional holographic analysis as well as dedicated
collider studies.

\begin{acknowledgments}
We thank Josh Erlich for valuable discussions. This work was supported by the
NSF under Grant PHY-0757481.
\end{acknowledgments}



\begin{thebibliography}{99}

\bibitem{technicolor}
S.~Weinberg,
Phys.\ Rev.\ D {\bf 19}, 1277 (1979);
E.~Farhi and L.~Susskind,
Phys.\ Rev.\ D {\bf 20}, 3404 (1979).

\bibitem{PT}
M.~E.~Peskin and T.~Takeuchi,
Phys.\ Rev.\ Lett.\  {\bf 65}, 964 (1990);
M.~E.~Peskin and T.~Takeuchi,
Phys.\ Rev.\ D {\bf 46}, 381 (1992).

\bibitem{etc}
S.~Dimopoulos and L.~Susskind,
Nucl.\ Phys.\ B {\bf 155}, 237 (1979).

\bibitem{hong}
  D.~K.~Hong and H.~U.~Yee,
  Phys.\ Rev.\  D {\bf 74}, 015011 (2006)
  [arXiv:hep-ph/0602177].

\bibitem{sanz}
  J.~Hirn and V.~Sanz,
  Phys.\ Rev.\ Lett.\  {\bf 97}, 121803 (2006)
  [arXiv:hep-ph/0606086].

\bibitem{cet1}
  C.~D.~Carone, J.~Erlich and J.~A.~Tan,
  Phys.\ Rev.\  D {\bf 75}, 075005 (2007)
  [arXiv:hep-ph/0612242].

\bibitem{acgr}
  K.~Agashe, C.~Csaki, C.~Grojean and M.~Reece,
  JHEP {\bf 0712}, 003 (2007)
  [arXiv:0704.1821 [hep-ph]].

\bibitem{cesh}
  C.~D.~Carone, J.~Erlich and M.~Sher,
  Phys.\ Rev.\  D {\bf 76}, 015015 (2007)
  [arXiv:0704.3084 [hep-th]].

\bibitem{hirayama}
  T.~Hirayama and K.~Yoshioka,
  JHEP {\bf 0710}, 002 (2007)
  [arXiv:0705.3533 [hep-ph]].

\bibitem{haba}
  K.~Haba, S.~Matsuzaki and K.~Yamawaki,
  Prog.\ Theor.\ Phys.\  {\bf 120}, 691 (2008)
  [arXiv:0804.3668 [hep-ph]].


\bibitem{piai}
  M.~Fabbrichesi, M.~Piai and L.~Vecchi,
  Phys.\ Rev.\  D {\bf 78}, 045009 (2008)
  [arXiv:0804.0124 [hep-ph]].

\bibitem{dandk}
D.~D.~Dietrich and C.~Kouvaris,
  Phys.\ Rev.\  D {\bf 78}, 055005 (2008)
  [arXiv:0805.1503 [hep-ph]];
  Phys.\ Rev.\  D {\bf 79}, 075004 (2009)
  [arXiv:0809.1324 [hep-ph]];
D.~D.~Dietrich, M.~Jarvinen and C.~Kouvaris,
  arXiv:0908.4357 [hep-ph].

\bibitem{mint}
  O.~Mintakevich and J.~Sonnenschein,
  JHEP {\bf 0907}, 032 (2009)
  [arXiv:0905.3284 [hep-th]].

\bibitem{kit}
  N.~Kitazawa,
  arXiv:0908.2663 [hep-th].

\bibitem{round}
M.~Round,
  arXiv:1003.2933 [hep-ph].

\bibitem{belitsky}
  A.~V.~Belitsky,
  arXiv:1003.0062 [hep-ph].

\bibitem{supertc}
  M.~Antola, S.~Di Chiara, F.~Sannino and K.~Tuominen,
  arXiv:1001.2040 [hep-ph];
M.~Jarvinen and F.~Sannino,
  arXiv:0911.2462 [hep-ph];
F.~Sannino,
  arXiv:0911.0931 [hep-ph].

\bibitem{luty}
  J.~A.~Evans, J.~Galloway, M.~A.~Luty and R.~A.~Tacchi,
  arXiv:1001.1361 [hep-ph].


\bibitem{also}
See also, K.~Kainulainen, K.~Tuominen and J.~Virkajarvi,
  arXiv:1001.4936 [astro-ph.CO].



\bibitem{adscft}
 J.~M.~Maldacena,
  Adv.\ Theor.\ Math.\ Phys.\  {\bf 2}, 231 (1998)
  [Int.\ J.\ Theor.\ Phys.\  {\bf 38}, 1113 (1999)]
  [arXiv:hep-th/9711200];
 S.~S.~Gubser, I.~R.~Klebanov and A.~M.~Polyakov,
  Phys.\ Lett.\ B {\bf 428}, 105 (1998)
  [arXiv:hep-th/9802109];
 E.~Witten,
  Adv.\ Theor.\ Math.\ Phys.\  {\bf 2}, 253 (1998)
  [arXiv:hep-th/9802150].

\bibitem{adsqcd}
 J.~Erlich, E.~Katz, D.~T.~Son and M.~A.~Stephanov,
  Phys.\ Rev.\ Lett.\  {\bf 95}, 261602 (2005)
  [arXiv:hep-ph/0501128];
 L.~Da Rold and A.~Pomarol,
  Nucl.\ Phys.\ B {\bf 721}, 79 (2005)
  [arXiv:hep-ph/0501218].

\bibitem{eandw}
J.~Erlich and C.~Westenberger,
Phys.\ Rev.\  D {\bf 79}, 066014 (2009)
[arXiv:0812.5105 [hep-ph]].

\bibitem{setc}
R.~S.~Chivukula, A.~G.~Cohen and K.~D.~Lane,
Nucl.\ Phys.\ B {\bf 343}, 554 (1990).

\bibitem{bt1}
A.~Kagan and S.~Samuel,
  Phys.\ Lett.\ B {\bf 252}, 605 (1990);
  Phys.\ Lett.\ B {\bf 270}, 37 (1991);
  Phys.\ Lett.\ B {\bf 284}, 289 (1992).

\bibitem{bt2}
E.~H.~Simmons,
Nucl.\ Phys.\ B {\bf 312}, 253 (1989).

\bibitem{bt3}
C.~D.~Carone and H.~Georgi,
Phys.\ Rev.\ D {\bf 49}, 1427 (1994)
[arXiv:hep-ph/9308205].

\bibitem{bt4}
C.~D.~Carone and E.~H.~Simmons,
Nucl.\ Phys.\ B {\bf 397}, 591 (1993)
[arXiv:hep-ph/9207273].

\bibitem{bt5}
C.~D.~Carone and M.~Golden,
Phys.\ Rev.\ D {\bf 49}, 6211 (1994)
[arXiv:hep-ph/9312303].

\bibitem{bt6}
C.~D.~Carone, E.~H.~Simmons and Y.~Su,
Phys.\ Lett.\ B {\bf 344}, 287 (1995)
[arXiv:hep-ph/9410242].

\bibitem{bt7}
V.~Hemmige and E.~H.~Simmons,
  Phys.\ Lett.\ B {\bf 518}, 72 (2001)
  [arXiv:hep-ph/0107117].

\bibitem{bt8}
  C.~X.~Yue, Y.~P.~Kuang, G.~R.~Lu and L.~D.~Wan,
  Mod.\ Phys.\ Lett.\ A {\bf 11}, 289 (1996);
  X.~L.~Wang, B.~Huang, G.~R.~Lu, Y.~D.~Yang and H.~B.~Li,
  Commun.\ Theor.\ Phys.\  {\bf 27}, 325 (1997);
 Y.~G.~Cao and Z.~K.~Jiao,
  Commun.\ Theor.\ Phys.\  {\bf 38}, 47 (2002).

\bibitem{bt9}
  M.~Antola, M.~Heikinheimo, F.~Sannino and K.~Tuominen,
  arXiv:0910.3681 [hep-ph].

\bibitem{Manohar:1983md}
  A.~Manohar and H.~Georgi,
  Nucl.\ Phys.\  B {\bf 234}, 189 (1984);
  H.~Georgi and L.~Randall,
  Nucl.\ Phys.\  B {\bf 276}, 241 (1986).

\bibitem{Amsler:2008zzb}
  C.~Amsler {\it et al.}  [Particle Data Group],
  Phys.\ Lett.\  B {\bf 667}, 1 (2008).

\bibitem{acp}
  K.~Agashe, R.~Contino and A.~Pomarol,
  Nucl.\ Phys.\  B {\bf 719}, 165 (2005)
  [arXiv:hep-ph/0412089].

\bibitem{Lee:1977eg}
  B.~W.~Lee, C.~Quigg and H.~B.~Thacker,
  Phys.\ Rev.\  D {\bf 16}, 1519 (1977).

\bibitem{hhg}
J.~F.~Gunion, H.~E.~Haber, G.~L.~Kane and S.~Dawson,
``The Higgs Hunter's Guide,''
{\it Redwood City, California: Addison-Wesley (1990) 404p.}

\bibitem{Barate:2003sz}
  R.~Barate {\it et al.}  [LEP Working Group for Higgs boson searches and
                  ALEPH Collaboration and  and],
  Phys.\ Lett.\  B {\bf 565}, 61 (2003)
  [arXiv:hep-ex/0306033].

\bibitem{ss}
  T.~Sakai and S.~Sugimoto,
  Prog.\ Theor.\ Phys.\  {\bf 114}, 1083 (2005)
  [arXiv:hep-th/0507073].



\end{thebibliography}
\end{document}